\documentclass{PoS}
\usepackage{graphicx}
\usepackage{caption}
\usepackage{subcaption}
\usepackage{amsmath,amsfonts,amssymb}

\let\OLDthebibliography\thebibliography
\renewcommand\thebibliography[1]{
\OLDthebibliography{#1}
\setlength{\parskip}{0pt}
\setlength{\itemsep}{0pt plus 0.3ex}
}
\usepackage{nicefrac}
\usepackage{graphicx,wrapfig,lipsum}

\title{Evaluation of expected solar flare neutrino events in the IceCube observatory}

\ShortTitle{Evaluation of expected solar flare neutrino events in the IceCube observatory}

\author{\speaker{G. de Wasseige}\\
       Interuniversity Institute for High Energies, Vrije Universiteit Brussel, Brussels, Belgium\\
        E-mail: \email{gdewasse@vub.ac.be}}

\author{P. Evenson\\
        Department of Physics and Astronomy, University of Delaware, Newark, DE, USA\\
        }
\author{K. Hanson\\
       WIPAC, University of Wisconsin-Madison, Madison, WISC, USA\\
        }
\author{N. van Eijndhoven\\
        Interuniversity Institute for High Energies, Vrije Universiteit Brussel, Brussels, Belgium\\
        }
\author{K.-L. Klein\\
       LESIA, Observatoire de Paris, Meudon, France\\
        }

\abstract{Since the end of the eighties and in response to a reported increase in the total neutrino flux in the Homestake experiment in coincidence with a solar flare, solar neutrino detectors have searched for solar flare signals. Neutrinos from the decay of mesons, which are themselves produced in collisions of accelerated protons with the solar atmosphere, would provide a novel window on the underlying physics of the acceleration process. 
For our studies we focus on the IceCube Neutrino Observatory, a cubic kilometer neutrino detector located at the geographical South Pole. Due to its Supernova data acquisition system and its DeepCore component, dedicated to low energy neutrinos, IceCube may be sensitive to solar flare neutrinos and thus permit either a measurement of the signal or the establishment of more stringent upper limits on the solar flare neutrino flux. We present an approach for a time profile analysis based on a stacking method and an evaluation of a possible solar flare signal in IceCube using the Geant4 toolkit.}

\FullConference{The 34th International Cosmic Ray Conference,\\
		30 July- 6 August, 2015\\
		The Hague, The Netherlands}

\begin{document}


\section{Introduction}
\paragraph{} In 1988, the Homestake experiment observed an increase in the total number of events in a possible correlation with energetic solar flares~\cite{homestake}. Bahcall predicted that if this increase were indeed due to solar flares,
this would lead to large characteristic signals in neutrino detection experiments~\cite{bahcall}. After Bahcall's prediction, neutrino detectors such as Kamiokande~\cite{kamiokande} and SNO~\cite{sno} performed several measurements. Even though these detectors used different solar flare samples and analyses, none of them has been able to confirm
the possible signal seen by Homestake. 
\paragraph{} Solar flares convert magnetic energy into plasma heating and kinetic energy of charged particles such as protons. Protons injected  downwards from the coronal acceleration region will interact with the dense plasma in the low solar atmosphere.
Because of the pion production channel through proton-nucleus interactions in the chromosphere and the subsequent decay of charged pions as presented in Equation~\ref{reaction}, solar flare neutrinos will be of great interest for investigating hadronic acceleration and the subsequent interactions in the chromosphere.
These neutrinos would indeed constitute a new way to study particle acceleration in solar flares, providing new constraints on e.g. the proton spectral index or the composition of the accelerated flux. Also, as pointed out by R.J. Murphy et al. in~\cite{interest} concerning gamma-rays and neutrons, solar flare neutrinos would offer the potential to learn about the structure and evolution of the flare environment.
\[p\, + \, p \quad \text{or} \quad p\, + \, \alpha 
\longrightarrow 
\left\{
\begin{array}{l}
 \pi^+ + X \\
  \pi^0 + X \\
    \pi^- + X \\
    \end{array}
\right.
\]
\begin{equation}
\label{reaction}
\end{equation}
\[\begin{array}{l}
\pi^+ \longrightarrow \mu^+ + \nu_{\mu} \qquad\qquad\qquad \mu^+ \longrightarrow e^+ + \nu_e + \bar\nu_{\mu} \\
\\
\pi^0 \longrightarrow 2 \gamma \\
\\
\pi^- \longrightarrow \mu^- + \bar{\nu}_{\mu} \qquad\qquad\qquad \mu^- \longrightarrow e^- + \bar{\nu}_e + \nu_{\mu} \\
\end{array}
\]
\paragraph{}Besides neutrinos, solar flares emit radiation across the entire electromagnetic spectrum~\cite{hudson}.
Gamma-rays are produced in this context by both neutral pion-decay and 
secondary electron $ $ bremsstrahlung. The pion-decay gamma-rays will be particularly useful in the neutrino search as outlined hereafter.
\section{Evaluation of the solar flare neutrino flux}
\subsection{Details and parameters of the simulation} \label{simulation}
\begin{figure}
        \centering
        \begin{subfigure}[b]{0.53\textwidth}
                \includegraphics[width=\textwidth]{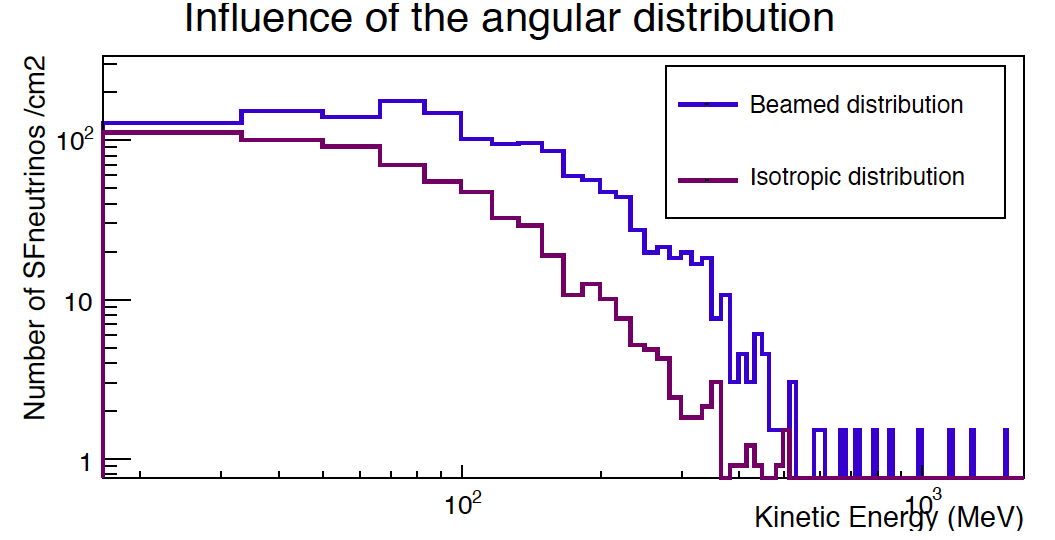}
                \caption{Angular distribution}
                \label{angular_distribution}
        \end{subfigure}
        \begin{subfigure}[b]{0.52\textwidth}
                \includegraphics[width=\textwidth]{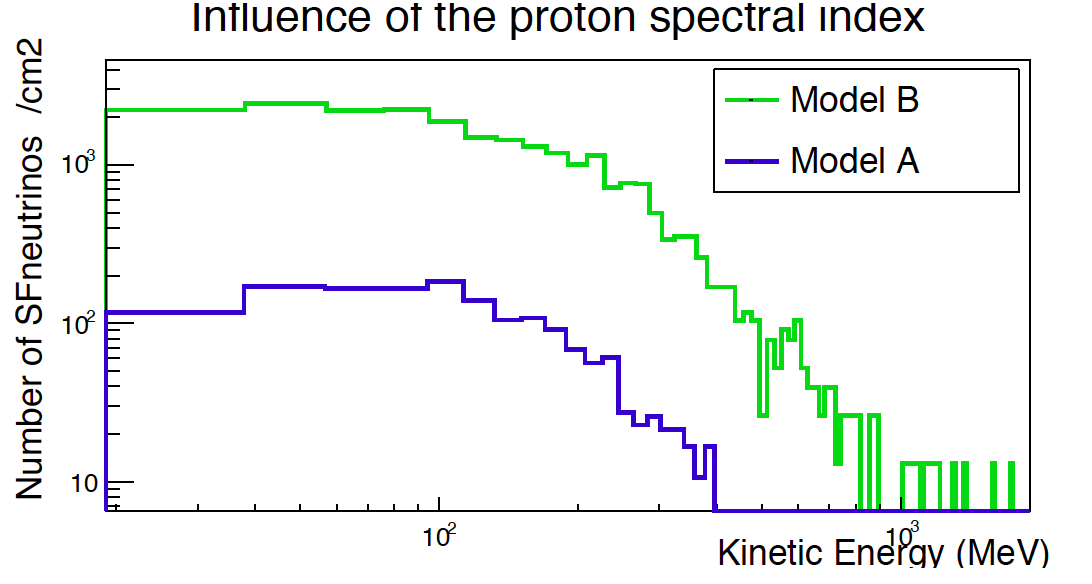}
                \caption{Spectral index}
                \label{spectralindex}
        \end{subfigure}%
        \begin{subfigure}[b]{0.53\textwidth}
                \includegraphics[width=\textwidth]{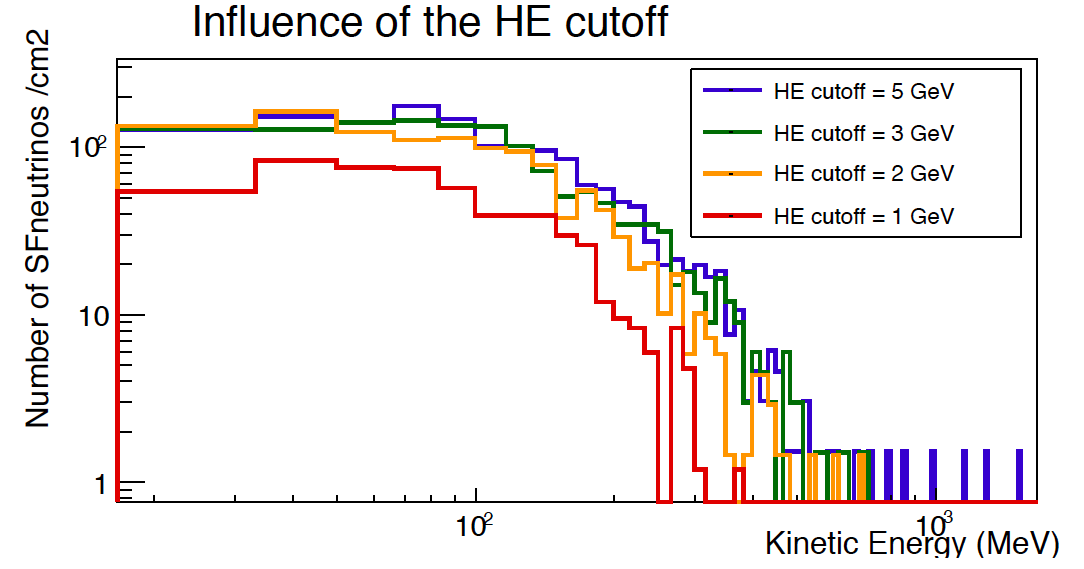}
                \caption{Upper cutoff}
                \label{hecutoff}
        \end{subfigure}
               \caption{Influence of specific parameters of the accelerated proton flux on the solar flare neutrino flux directed to the Earth. The studied parameter is mentioned below each plot. For the influence of the spectral index, a \textit{beamed distribution} has been assumed as well as for the upper cutoff for which the \textit{Model A} has been used.}\label{parameter}
\end{figure}

In order to evaluate the neutrino flux produced by a single solar flare, we have designed a Geant4 simulation~\cite{geant4} of proton-nucleus interactions in the chromosphere. We have used the "Model of Chromospheric Flare Regions"~\cite{machado} to define the density profile of the interaction region, i.e. the chromosphere~\cite{ramaty}. We did not simulate the hadron acceleration nor any other magnetic effects in the current simulation. Because we expect these non-simulated effects to have a rather important impact on the neutrino flux produced in the direction of the Earth, we have simulated the two most extreme cases - the most realistic case expected to be somewhere in between - which are:
\begin{itemize}
\item a proton beam tangent to the chromosphere in the direction of the Earth (henceforth referred to as  \textit{beamed distribution})
\item an isotropic distribution of protons injected vertically into the chromosphere (i.e. \textit{isotropic distribution})
\end{itemize}
Both of these proton distributions have been injected from the corona in the direction of the chromosphere. The influence of these two distributions on the solar flare neutrino flux directed to the Earth is shown in Figure~\ref{angular_distribution} where the effect of the beaming yielding a harder spectrum is clearly visible.
\paragraph{}
In order to be as model-independent as possible, one of the proton spectra injected in this simulation has been derived from gamma-ray observations of the June 3, 1982 event by R.J.~Murphy and R.~Ramaty~\cite{murphyandramaty}. We will call this $E^{-3.1}$ spectral index \textit{Model A}.
A second proton spectrum, a generic $E^{-2}$, has also been simulated and will be mentioned as \textit{Model B}.
The details of these two different models are presented in Table \ref{table-models} and their influence on the solar flare neutrino flux in direction of the Earth is shown in Figure~\ref{spectralindex}.  Here it is seen that the $E^{-2}$ spectrum yields an order of magnitude more neutrinos than model A, and is probably a very optimistic scenario.
\begin{table}[h!]
\begin{center}
\begin{tabular}{c| c | c}
Model & Proton spectrum & Total number of protons \\ \hline
Model A & $E^{-3.1}$ &$2.2\times10^{33}$ \\ 
Model B & $E^{-2}$ &$7.9\times10^{32}$
\end{tabular}
\end{center}
\caption{\label{table-models} Details of the two different models of proton spectra simulated. The \textit{Total number of protons} represents the total number of protons which have been accelerated to an energy E $>$ 30~MeV by the solar flare. These numbers have been obtained by considering that the total solar flare energy is the same in both models~\protect\footnotemark .}
\end{table}

\footnotetext{Even though this approximation  may not be correct for a general flare, it becomes reasonable for the solar flare sample which will be considered (see Section~\ref{novel idea}).} 

The production of neutrinos by protons with energies below 500 MeV can be neglected. We furthermore considered that no protons were accelerated beyond some energy, the upper cutoff, which was treated as a free parameter.  
The evidence of an upper cutoff in the accelerated proton spectrum has been demonstrated by Dj. Heristchi et al.~\cite{trottet}. As highlighted in the same paper~\cite{trottet}, the upper cutoff may not be as sharp as they have defined it but may be represented by a rapid change in the spectral index at high energies. Considering that the latter would produce more neutrinos at high energies than the former one, we have not simulated this effect in the present simulation~\footnote{A future version of the simulation will study the impact of this potential contribution of a softer spectrum above the upper cutoff.}. 
Instead, we have performed our simulations for different upper cutoff values, i.e. 1, 2, 3 and 5~GeV. The difference in the solar flare neutrino flux directed to the Earth due to the variation of the upper cutoff of the accelerated proton distribution is shown in Figure~\ref{hecutoff}. Note that the cutoff values above or equal to 3~GeV basically yield the same result.

The composition of the accelerated flux is not investigated here; a pure accelerated proton flux has been assumed~\footnote{The composition of the simulated atmosphere is nevertheless a mixing of Hydrogen and Helium in a ratio He:H = 1:9.}. A mixed accelerated flux would in any case boost the neutrino flux directed to the Earth.

\subsection{Solar flare neutrino flux}
In order to evaluate the feasibility that IceCube would be able to see a signal from solar flares, we have investigated the tangent injection of a proton spectrum based on the \textit{Model A} standing here as the most optimistic but realistic case. The solar flare neutrino flux expected for the two extreme upper cutoff values considered in this simulation are presented in Table~\ref{table-flux}. One will note that the spectrum has been divided into two parts : $< 100$~MeV  and $> 100$~MeV. Even though this distinction will be convenient in Section~\ref{icecube}, it follows, as expected, a physical reality of two different behaviors of the neutrino spectrum as shown in Figures~\ref{parameter}; the low energy part is indeed due to the bulk of the pion-decay bump while the higher energies represent the tail of this bump.

\begin{table}[t]
\begin{center}
\begin{tabular}{c |  c | c}
  \hline
Energy range &Neutrino fluence at Earth &Spectrum\\
  \hline
10 - 100~MeV & 398$^a$ - 770$^b$ $\nu$ cm$^{-2}$& E$^0$ \\
100 - <1000~MeV & 221$^a$ - 783$^b$ $\nu$ cm$^{-2}$ & E$^{-2.3}$ \\
  \hline
\end{tabular} 
\end{center}
\caption{\label{table-flux} Expected neutrino fluence at Earth and spectrum for one single solar flare. The numbers presented in this table have been obtained from the simulation of a tangent injection of the proton spectrum described by the \textit{Model A} with an upper cutoff at 1~GeV (a) and 5~GeV (b).}
\end{table}

\begin{figure}[h!]
    \centering
    \includegraphics[width=0.7\textwidth]{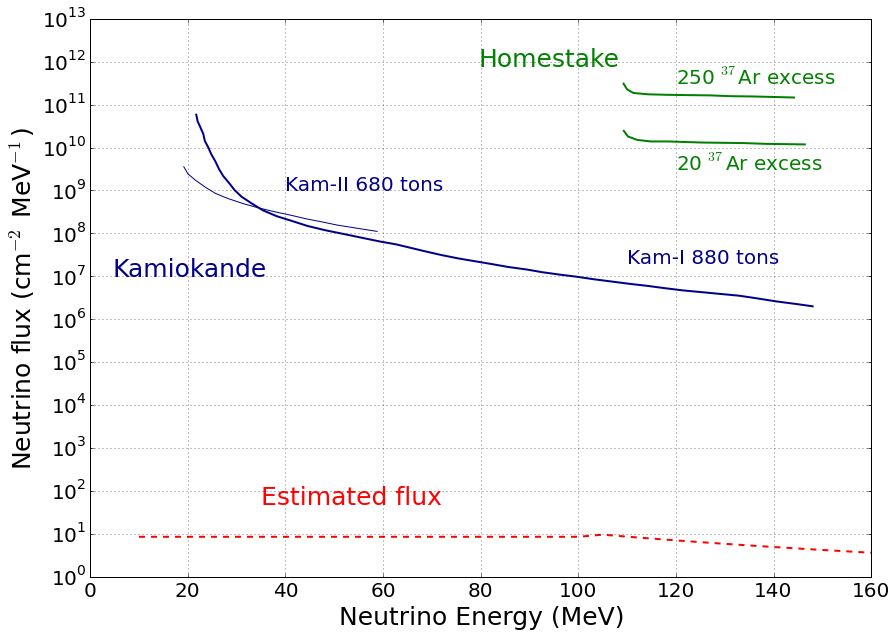}
    \caption{ Comparison of the single solar flare neutrino flux obtained from the simulation with the limits set up by  Homestake~\cite{homestake} and Kamiokande~\cite{kamiokande}. These limits have been obtained interpreting the sensitivity plot in~\cite{kamiokande} as neutrino flux (cm$^{-2}$ MeV$^{-1}$) versus neutrino energy (MeV) for the ($\nu_e:\bar\nu_e:\nu_{\mu}:\bar\nu_{\mu} = 1:1:2:2$) case. \label{comp}}
\end{figure}

In order to evaluate how reasonable our simulated single solar flare neutrino flux is, we have compared it with the limits set by Homestake~\cite{homestake} and Kamiokande~\cite{kamiokande}. One can easily see in Figure~\ref{comp} that neither Kamiokande nor Homestake were supposed to see a signal from a single solar flare considering the parameters which have been used to design the present simulation. Nevertheless, some dramatically energetic solar flares such as the "X-Whatever-flare" (X28) from November 4, 2003 might produce a larger neutrino flux. The solar flare neutrino flux obtained through the simulation described in the present work is therefore consistent with previous observations.
\section{IceCube and solar flare neutrinos} \label{icecube}
IceCube, the neutrino observatory buried in the Antarctic ice, is made of 86 vertical strings, each one containing 60 optical
modules connected to it. These 5160 modules, located between 1450 and 2450 meters below the South Pole surface, detect Cherenkov light emitted by charged leptons produced in neutrino interactions with the nearby ice or underlying bedrock. Even though IceCube has been designed to detect TeV neutrinos~\cite{hese}, it contains a dense sub-detector and a data acquisition system dedicated to study neutrinos with lower energies.
\begin{itemize}
\item[]\textsc{DeepCore (DC)}: This sub-detector of IceCube is made of 15 strings (7 IceCube + 8 additional). DeepCore is able to detect neutrinos of 100~MeV and above because of its higher density of optical modules. This denser part allows IceCube to be more sensitive to atmospheric neutrino oscillation studies and dark matter searches ~\cite{deepcore}.
\item[]\textsc{Supernova data acquisition system (I3/SNDAQ)}: This system has been implemented in view of the detection of supernova neutrinos.  By searching for an increase in the noise level of the detector, the Supernova data acquisition system is able to detect fast transient bursts of neutrinos with an average neutrino energy as low as 10~MeV~\cite{sn}. It could therefore be used to detect large neutrino bursts in general.
\end{itemize}
In order to evaluate the sensitivity of IceCube to solar flare neutrinos, a Monte-Carlo simulation of the production and propagation of the Cherenkov light related to the solar flare neutrino interactions in the nearby ice has been carried out. The effective mass M$_{\text{eff}}$ (reflecting the fiducial volume) is established this way. Therefore, the number of events detected by IceCube is given by:\[\]
\begin{equation} N_{\text{events}}=  \int_{E_{\text{min}}}^{E_{\text{max}}}N_A \sigma(E) M_{\text{eff}}(E) \frac{d\phi(E)}{dE}dE \end{equation}\[\]
Where $N_A$ stands for the Avogadro number, $\sigma(E)$ is the neutrino-ice cross-section at the considered energy \cite{strumia,oxygene,fargion} and $ \nicefrac{d\phi(E)}{dE}$ the solar flare neutrino spectrum expressed in Table~\ref{table-flux}.
This number of events has been calculated for both DeepCore and the Supernova system as shown in Table~\ref{table-events}~\footnote{The oscillation mixing is not taken into account in the present calculation. The effect is expected to be minor since IceCube is sensitive to all electron and muon channels.}.
\\
In order to estimate if the computed numbers of events would lead to a significant detection, we have evaluated the expected background integrated over the duration of a solar flare. The number of background events presented in Table~\ref{table-events} has been obtained for an integration time of 4 minutes (the reason is detailed in Section~\ref{novel idea}). Table~\ref{table-events} shows the required number of events for a 3~$\sigma$ detection.

\begin{table}[t]
\begin{center}
\begin{tabular}{c| c |  c | c}
  \hline
Detector & Energy range & Expected number of events & Required for  \\
 &  & for one solar flare&  a 3$\sigma$ detection \\
  \hline
I3/SNDAQ& 10 - 100~MeV & 100$^a$ - 193$^b$&  50000 \\
DC& 100 - <1000~MeV & 25$^a$ - 89$^b$ & 100 \\
  \hline
\end{tabular}

\end{center}
\caption{\label{table-events} Comparison of the expected number of events for one single solar flare in the Supernova system (I3/SNDAQ) and DeepCore (DC) with the required number of events for a 3$\sigma$ detection. The numbers have been derived from results presented in Table~\protect\ref{table-flux} for a upper cutoff at 1~GeV (a) and 5~GeV (b). }
\end{table}

One can easily read from Table~\ref{table-events} that a 3~$\sigma$ detection does not seem possible using only the Supernova  data acquisition system. On the other hand, DeepCore appears to be a promising way to detect solar flare neutrinos with an energy above 100~MeV.


\section{Novel idea for a stacking analysis} \label{novel idea}
\begin{wrapfigure}{t!rh!}{6.5cm}
\includegraphics[width=0.45\textwidth]{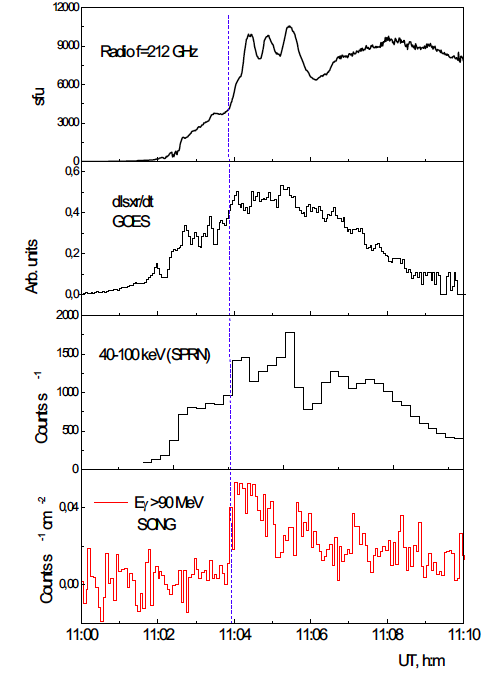}
    \caption{Time series of the impulsive phase of a pion-flare for the October 28, 2003 event \cite{icrc2011}. \label{gamma_idea}}\end{wrapfigure}
\begin{figure}[th!]
    \centering
    \includegraphics[width=0.7\textwidth]{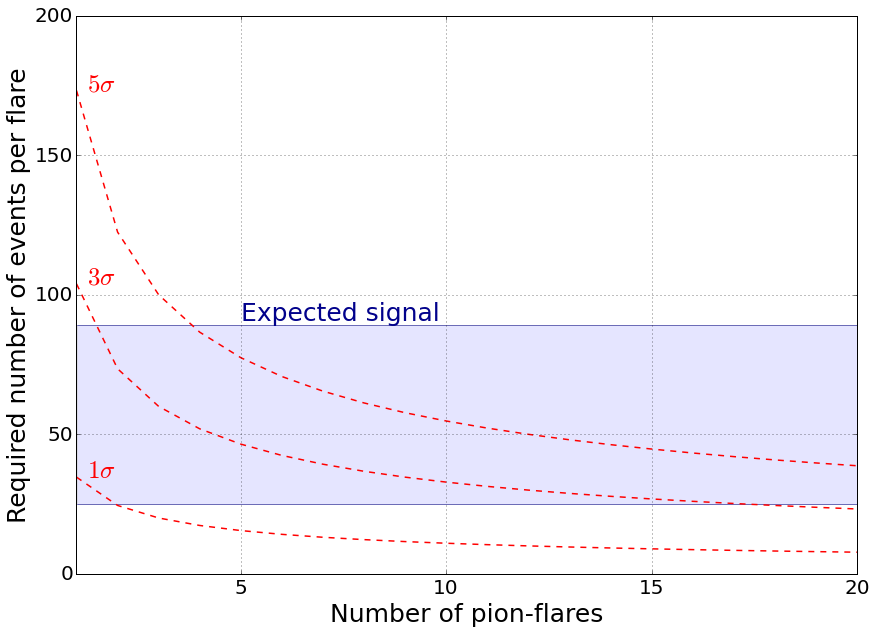}
    \caption{Comparison of the expected signal and the required number of events per flare for a X~$\sigma$ detection in DeepCore. The expected signal bounds have been obtained from Table~\protect\ref{table-events}. One can easily see that a 5~$\sigma$ detection would be obtained with a sample of only 5 pion-flares in the case of Model A, using a tangent injection and an upper cutoff of 5 GeV (see Section~\protect\ref{simulation}).\label{bkgvssignal}}
\end{figure}

As shown in Table~\ref{table-events}, there is  a realistic chance to detect solar flare neutrinos above 100~MeV with DeepCore.
However, even if a detection during an individual flare might be possible, a stacking analysis will be more promising in any case.
In order to maximize the chance of a detection, we will define new criteria on the solar flare selection rather than considering systematically every solar flare with a total number of registered photon
counts $I >I_{threshold}$ as it has been done in~\cite{sno} for example.
The sample will be restricted to solar flares for which pion-decay gamma-rays have been observed by e.g. Fermi-LAT~\cite{fermi-pion-flare}. We will call these specific solar flares "pion-flares".
Moreover, considering that the background in IceCube described in Section~\ref{icecube} will increase with the integration time of the analysis, we will focus our search on the impulsive phase of the pion-flares. Indeed, the pion-decay gamma-ray emission in this impulsive phase of the flare is a burst with a full-width at half-maximum of maximum 4 minutes as illustrated in Figure~\ref{gamma_idea}~\cite{icrc2011}. 
Using satellite data as a reference, a time-profile analysis (such as in~\cite{nick}) of a few solar flares would lead to a detection in the DeepCore region of IceCube assuming the current model (see Section \ref{simulation}).
Considering that a minimum of 5 pion-flares would be required for a $5~\sigma$ detection as shown in Figure~\ref{bkgvssignal}, using the list suggested by the Fermi collaboration~\cite{fermi-pion-flare} would constitute an excellent starting point.


\section{Summary}  Considering the current parameters related to the hadron acceleration in solar flares, we have designed a Geant4 simulation in order to evaluate the solar flare neutrino flux directed to the Earth. The validity of our calculated neutrino spectrum compared to previous limits has been demonstrated. Evaluating the IceCube sensitivity to these specific neutrinos, we have shown that there is a realistic chance to see a signal from a single solar flare by detecting neutrinos above 100~MeV. Nevertheless, in view of maximizing the probability of a detection, we have developed a new way to search for solar flare neutrinos by defining a specific solar flare sample as well as a narrow time-window in which we expect the neutrino production. According to the current simulation,  five specific solar flares in a stacked analysis would be enough for a 5~$\sigma$ detection in the most optimistic case but for a realistic proton spectrum.

\end{document}